\documentclass[prd,twocolumn,amsmath,amssymb,nofootinbib,preprintnumbers,balancelastpage]{revtex4}
\usepackage{graphicx}
\usepackage{hyperref}
\hypersetup{
    pdfnewwindow=true,      
    colorlinks=true,        
    linkcolor=black,        
    citecolor=blue,         
    filecolor=blue,         
    urlcolor=blue           
}
\begin{document}
\title{\Large {\bf{The Higgs Mass and the Stueckelberg Mechanism in Supersymmetry }}}
\author{Pavel Fileviez P\'erez}
\affiliation{\vspace{0.15cm} \\  Particle and Astro-Particle Physics Division \\
Max-Planck Institute for Nuclear Physics {\rm{(MPIK)}} \\
Saupfercheckweg 1, 69117 Heidelberg, Germany}
\author{Sogee Spinner}
\affiliation{\vspace{0.15cm} \\ University of Pennsylvania, Department of Physics and Astronomy \\
209 South 33rd Street, Philadelphia, PA 19104-6396, USA}

\begin{abstract}
We investigate a class of theories where the mass of the lightest Higgs boson of the Minimal Supersymmetric Standard Model 
(MSSM) can be larger than the Z gauge boson mass at tree level. In this context the MSSM fields feel a new force, whose corresponding gauge boson attains its mass through the Stueckelberg mechanism. We show how one can 
achieve a Higgs mass around 126 GeV without assuming a heavy stop spectrum or a large stop
trilinear term. The application of this class of models to the conservation of R-parity is also discussed.
\end{abstract}

\maketitle

\section{Introduction}
New experimental results from the Large Hadron Collider have questioned the existence of new physics relevant to the weak scale.
In particular, the discovery of a neutral scalar with properties very similar to those of the standard model (SM) Higgs boson and a mass $m_h \sim 126$ GeV, has set strong constraints on the spectrum of physics beyond the SM, especially the 
Minimal Supersymmetric Standard Model (MSSM). 

The friction between the Higgs boson discovery and the MSSM originates from the fact that the MSSM predicts an upper bound on the lightest Higgs mass at tree level. This bound is simply the $Z$ gauge boson mass, $91$ GeV. Large quantum corrections are then required in order to achieve a Higgs mass in agreement with the experimental measurements. These corrections require heavy stops or a large trilinear stop term, with somewhat lighter stops. See Ref.~\cite{review} for a recent review of the Higgs mass in supersymmetric models.

There are two scenarios beyond the MSSM, which allow a Higgs mass larger than the $Z$ mass at tree level:
\begin{itemize}

\item $F$-term contributions: In this case, the MSSM Higgs interact with new fields which contribute to the Higgs quartic coupling through $F$-terms.  There are three such possible sets of fields: 

\begin{itemize}

\item The singlet, $\hat{S} \sim (1,1,0)$, which couples as $\hat{S} \hat{H}_u \hat{H}_d$. 
This scenario is referred to as the NMSSM or next-to-minimal supersymmetric SM. 
See Ref.~\cite{NMSSM} for details.

\item The hyperchargeless triplet, $\hat{\Sigma} \sim (1,3,0)$, which also modifes the Higgs couplings through the new interaction, $ \hat{H}_u  \hat{\Sigma} \hat{H}_d$. For details see Refs.~\cite{Espinosa:1991wt,Espinosa:1991gr,Sogee1, Basak, DiChiara}.

\item Two $Y=\pm1$ triplets, $\hat{\Delta} \sim (1,3,1)$ and $\hat{\bar{\Delta}} \sim (1,3,-1)$, so that the superpotential now contains the new interactions $\hat{H}_u  \hat{\bar{\Delta}} \hat{H}_u$ and $\hat{H}_d  \hat{{\Delta}} \hat{H}_d$. Not only do these fields contribute to the tree level Higgs mass but they also generates neutrino masses through the term $\hat{L}  \hat{{\Delta}} \hat{L}$. This is referred to as the type II seesaw mechanism. For a recent discussion of this scenario see Ref.~\cite{Espinosa:1991gr,Sogee2}.

\end{itemize}

The NMSSM and the real triplet scenario only increase the Higgs mass when the ratio between the MSSM Higgs vacuum expectation values, $\tan \beta = v_u/v_d$, is small. On the other hand, the enhancement in the Type II scenario can occur for any value of $\tan \beta$ and 
provides a connection to neutrino masses. Unfortunately, the introduction of triplets spoils the unification of the SM gauge couplings at high scales, if one assumes the presence of the desert between the weak and unification scales. 

\item $D$-term contributions: The introduction of new gauge symmetries, Abelian or non-Abelian, relevant for the Higgs bosons of the MSSM, modify the quartic terms of the Higgs fields and can, therefore, increase the lightest Higgs boson mass at tree level. Such new symmetries require a new Higgs sector, which is used to break the new symmetry at some scale above the weak scale. This type of mechanism has also been investigated by many experts in the field. See for example Refs.~\cite{Batra,Medina} for details.

\end{itemize}
In this letter, a third, new, possibility is presented for increasing the Higgs mass at tree level beyond the Z mass. 
We introduce a new Abelian force which changes the MSSM Higgs fields interactions, as in the $D$-term case, but the mass of the new gauge boson is generated through the Stueckelberg mechanism~\cite{Stueckelberg,review1,SMSt}. This modifies the Higgs mass at tree level, while keeping all of the beneficial features of the MSSM. We investigate this model in detail, showing the dependence of the Higgs mass on the new parameters. Since the Stueckelberg mechanism leaves the new Abelian symmetry unbroken in the MSSM sector, one can use the same Abelian symmetry 
to explain the conservation of R-parity and solve a second major issue present in the MSSM. 
This is possible when the new gauge symmetry is related to $B-L$.
In this case, the new symmetry has a dual role: it can increase the Higgs mass 
and allow for the conservation of matter parity.

This article is organized as follows: In section II we discuss the Stueckelberg mechanism for a 
simple Abelian extension of the MSSM, while in section III we discuss the Higgs and neutral gauge boson spectrum. 
Finally, we discuss the main features of this scenario and its dark matter possibilities.
%
\section{The Stueckelberg Mechanism}
%
The simple extension of the MSSM investigated in this article is based on the gauge group 
\begin{equation}
	SU(3)_C \otimes SU(2)_L \otimes U(1)_Y \otimes U(1)_X,
\end{equation}
where $U(1)_X$ is a new Abelian symmetry, relevant for all the MSSM fields. 
The Stueckelberg extension of the MSSM has been studied in Ref.~\cite{Kors:2004ri} 
for the case where the new symmetry is only relevant for the hidden sector, while a
$B-L$ Stueckelberg extension was studied in~\cite{Feldman:2011ms}.

We proceed by leaving the $x$ charges of the MSSM particle content completely general, but
allow for the $\mu \hat{H}_u \hat{H}_d$ term in the superpotential, requiring 
the $x$ charge of $\hat{H}_u$ to be the opposite of $\hat{H}_d$,
\begin{align}
	\hat{H}_u & \sim (1,2,1/2, x_{H_u}), 
\end{align}
and
\begin{align} 
\hat{H}_d  \sim (1,2,-1/2, -x_{H_u}).
\end{align}
While symmetry breaking in the SM gauge group is defined by the Higgs mechanism, 
the mass of the new $U(1)_X$ gauge boson is generated by the Stueckelberg mechanism. 
In this case, the relevant Lagrangian for this discussion is given by
\begin{eqnarray}
{\cal L} &\supset& \frac{1}{4} \int d^2 \theta \ {\cal W}_X {\cal W}_X \ + \  \frac{1}{4} \int d^2 \bar{\theta}  \  \bar{{\cal W}}_X \bar{{\cal W}}_X \nonumber \\
& + & {\cal L}_\text{Higgs} \ + \ {\cal L}_\text{St},
\end{eqnarray}
where ${\cal W}_X = - 1/4 \bar{D} \bar{D} D \hat{X}$ is defined in the usual way, and $\hat{X}$ 
is the vector superfield associated with the new Abelian force. For simplicity, we neglect 
any possible Kinetic mixing between the Abelian symmetries.
The kinetic terms for the MSSM Higgs fields are given by
\begin{eqnarray}
{\cal L}_\text{Higgs} &=&
	\int d^2 \theta d^2 \bar{\theta} \
	\hat{H}_u^\dagger e^
	{
		g_2 \hat{W} + g_1\hat{B} + g_X x_{H_u} \hat{X}
	}\hat{H}_u 
	\nonumber \\
& + & 
	\int d^2 \theta d^2 \bar{\theta} \
	\hat{H}_d^\dagger
	e^
	{
		g_2 \hat{W} - g_1  \hat{B} - g_X x_{H_u} \hat{X}
	} \hat{H}_d,
\end{eqnarray}
where the $SU(2)_L$ and $U(1)_Y$ vector superfields are represented by $\hat{W} = \hat W^a \sigma^a$ 
($\sigma^a$ being the Pauli spin matrices) and $\hat{B}$, respectively.
Since the new gauge boson associated with $U(1)_X$ acquires mass through the Stueckelberg mechanism, 
an extra chiral superfield $\hat S$ is needed. In order to set our notation we define the vector and chiral superfields as 
\begin{eqnarray}
\hat{X} &=& - \theta \sigma^\mu \bar{\theta} X_\mu +  i  \theta \theta \bar{\theta} \bar{\tilde{X}} -  i  \bar{\theta} \bar{\theta} \theta \tilde{X} + \frac{1}{2} \theta \theta \bar{\theta} \bar{\theta} D_X, \\
\hat{S} &= & \frac{1}{2} (\rho + i \sigma) + \theta \tilde{S} + i \theta \sigma^\mu \bar{\theta} \frac{1}{2} (\partial_\mu \rho + i \partial_\mu \sigma) 
 + \theta \theta F_s \nonumber \\
 &+ & \frac{i}{2} \theta \theta \bar{\theta} \bar{\sigma}^\mu \partial^\mu \tilde{S} + \frac{1}{8} \theta \theta \bar{\theta} {\theta} ( \Box \rho + i \Box \sigma). 
\end{eqnarray}
 The Stueckelberg gauge transformations for $\hat X$ and $\hat S$ are
\begin{eqnarray}
	\hat{X} &\to& \hat{X} +  \hat{\Lambda} + \hat{\Lambda}^\dagger, \  \  {\text{and}} \  \ 
	\hat{S} \to  \hat{S} - M_X \hat{\Lambda}.
\end{eqnarray}
These transformations allow the following terms in the Lagrangian
\begin{eqnarray}
\label{eq:LSt}
{\cal L}_\text{St} &=& \int d^2 \theta d^2 \bar{\theta}  \left(  M_X  \hat{X}  +  \hat{S} +  {\hat{S}}^\dagger  \right)^2.
\end{eqnarray}
Notice that in this approach, only one chiral superfield, $\hat S$, is needed to generate mass for the Abelian gauge boson. This can be compared to a typical Higgs mechanism, which requires at least two chiral superfields, due to anomaly cancellation, when the chiral superfields have a traditional gauge charge. 

Neglecting kinetic terms, in component form, Eq.~(\ref{eq:LSt}) becomes
\begin{eqnarray}
\label{eq:L.St}
	{\cal L}_{St} &\supset& - \frac{1}{2} M_X^2 X_\mu X^\mu \ + \  M_X D_X \rho + 2 |F_s|^2 \nonumber \\
	&+ & i M_X\left(\tilde X \tilde S + \text{H.c.}\right),
\end{eqnarray} 
where the first term is the mass term for the new gauge boson and the last mixes the Stueckelbergino with the xino. The equation of motion for $D_X$ yields
\begin{equation}
	D_X = -M_X \rho - \frac{1}{2} g_X \sum_\phi x_\phi |\phi|^2,
\end{equation}
where $x_\phi$ is the $U(1)_X$ charge of $\phi$, a scalar field. Substituting this into Eq.~(\ref{eq:L.St}) and focusing only on the scalar potential yields
\begin{eqnarray}
\label{eq:L.St1}
	{\cal L}_{St} &\supset & -\frac{1}{2} M_X^2 \rho^2 - \frac{1}{2} g_X \, M_X \, \rho \, \sum_\phi \phi^* x_\phi \phi \nonumber \\
	&-& \frac{1}{8} g_X^2 \left(\sum_\phi \phi^* x_\phi \phi\right)^2.
\end{eqnarray}
The last term in Eq.~(\ref{eq:L.St1}) is the regular $D$-term, while the middle term is due to the Stueckelberg Mechanism. This middle term mixes $\rho$ with the MSSM Higgs fields once the latter fields acquire a vacuum expectation value (VEV). The combination of these two terms have important implications for the lightest Higgs mass. For more details on the application of the Stueckelberg mechanism to Abelian extensions of the MSSM, see Refs.~\cite{Kors:2004ri,Kors:2005uz} . See also Ref.~\cite{Feldman:2011ms} for the application of this mechanism to the conservation of $R$-parity when the new symmetry is $U(1)_{B-L}$. 
%
\section{The Lightest Higgs Boson Mass}
Using the above interactions, the scalar potential for the neutral Higgs fields is
 \begin{eqnarray}
V (H_u^0, H_d^0, \rho) &=& V_{MSSM} + V_X,  
\end{eqnarray}
with
 \begin{eqnarray}
V_{MSSM} &=& ( |\mu|^2 + m_{H_u}^2 ) |H_u^0|^2 +  ( |\mu|^2 + m_{H_d}^2 ) |H_d^0|^2 \nonumber \\
&- & (b H_u^0 H_d^0 + h.c.) + \frac{1}{8} (g_1^2+ g_2^2) (|H_u^0|^2 - |H_d^0|^2)^2, \nonumber \\
\\
 \label{eq:Vx}
V_X &=& \frac{1}{8} g_X^2 x_{H_u}^2 (|H_u^0|^2 - |H_d^0|^2)^2 + \frac{1}{2} ( M_X^2 +  \tilde{m}_\rho^2 ) \rho^2 \nonumber \\
&+ & \frac{1}{2} g_X \, x_{H_u} \, M_X \,  \rho \ (|H_u^0|^2 - |H_d^0|^2),
\end{eqnarray}
where $\tilde m_\rho^2$ is the soft mass of $\rho$. In our notation, the neutral MSSM fields are given by
 \begin{eqnarray}
H_u^0 &=& \frac{h_u + v_u}{\sqrt{2}} + i \frac{A_u}{\sqrt{2}},
\\
 H_d^0 &=&  \frac{h_d + v_d}{\sqrt{2}} + i \frac{A_d}{\sqrt{2}}.
 \end{eqnarray}
The last term in Eq.~(\ref{eq:Vx}) induces a VEV, $v_\rho$, for $\rho$:
\begin{equation}
	v_\rho = \frac{g_X M_X x_{H_u} \left(v_d^2 - v_u^2\right)}{4 \left( M_X^2 + \tilde m_\rho^2\right)}.
\end{equation}
As will be argued later, since the MSSM Higgs fields have non-zero $x$ charge, their VEVs will induce $X$-$Z$ mixing. Such mixing is highly constrained by electroweak precision tests and will, therefore, force $M_X^2 \gg M_Z^2$. This, in turn, means that $v_\rho^2 \ll v_u^2, v_d^2$. 

The MSSM minimization conditions read as
\begin{align}
	\frac{1}{2} M_{Z_0}^2 + \frac{x_{H_u}^2 g_X^2v^2 \tilde m_\rho^2}{8\left(M_X^2+\tilde m_\rho^2\right)}  & = 
	-|\mu|^2 + \frac{\tan^2 \beta \,  m_{H_u}^2 - m_{H_d}^2}{1 - \tan^2 \beta},
\\
	\frac{2 b}{\sin 2 \beta} & = m_{H_u}^2+m_{H_d}^2 + 2 |\mu|^2,
\end{align}
where $v^2 = v_u^2+v_d^2$ and $M_{Z_0}^2 = \left(g_1^2+g_2^2\right) v^2 /4$. 

In the basis, $\left(B_\mu,W^0_\mu, X_\mu\right)$, the neutral gauge boson mass matrix is
\begin{equation}
	\begin{pmatrix}
		\frac{1}{4} g_1^2 v^2
		&
		-\frac{1}{4} g_1 g_2 v^2
		&
		-\frac{1}{4} g_1 g_X x_{H_u} v^2
	\\
		-\frac{1}{4} g_1 g_2 v^2
		&
		\frac{1}{4} g_2^2 v^2	
		&
		-\frac{1}{4} g_2 g_X x_{H_u} v^2
	\\
		-\frac{1}{4} g_1 g_X x_{H_u} v^2
		&
		-\frac{1}{4} g_2 g_X x_{H_u} v^2
		&
		M_X^2 +\frac{1}{4} g_X^2 x_{H_u}^2 v^2
	\end{pmatrix}.
\end{equation}
\begin{figure}[t]
	\includegraphics[scale=0.53]{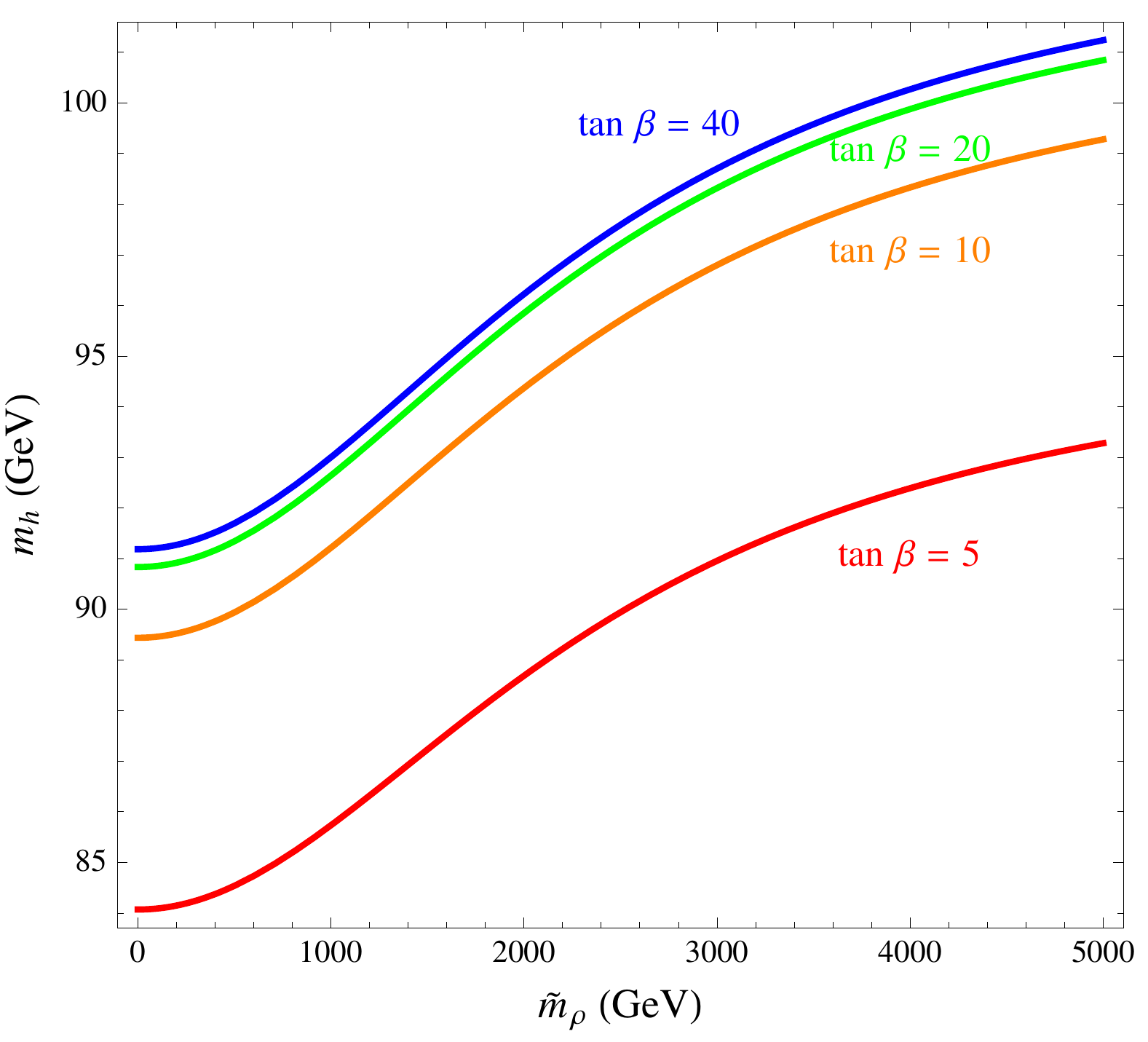}
	\caption{The tree level lightest Higgs mass, $m_h$, versus the soft 
	mass parameter for $\rho$, $\tilde m_\rho$. Curves are displayed for $\tan \beta = 5, \, 10, \, 20$ and 40 and assuming $x_{H_u} = 1$, $M_X=2.5$ TeV, $g_X = 0.4$ and $m_A = 500$ GeV.
	A significant modification, pushing the tree level Higgs mass above the $Z$ mass, 
	is only possible when $\tilde m_\rho$ is comparable to $M_X$.}
	\label{fig:mh0}
\end{figure}
Rotating away the zero mass photon and expanding in terms of $v^2/M_X^2$ yields the two massive eigenstates, $Z$ and $Z_X$ with masses:
\begin{eqnarray}
	M_Z^2 & \approx & \frac{1}{4} \left(g_1^2 + g_2^2 \right)v^2 \left(1 - \frac{1}{4} g_X^2 x_{H_u}^2 \frac{v^2}{M_X^2}\right),
	\\
	M_{Z_X}^2 & \approx & M_X^2 \left(1+ \frac{1}{4} g_X^2 x_{H_u}^2 \frac{v^2}{M_X^2}\right). \\ \nonumber
	\label{eq:Z.mass}
\end{eqnarray}
The most model independent bound on $M_X$ comes from the $\rho$ parameter, which constrains new contributions to the $Z$ mass. Assuming $g_X x_{H_u} = 0.4$, the $\rho$ parameters places the lower bound, $M_X > 1.5$ TeV, at two sigma. This justifies our earlier assumption that $M_X \gg v$.
As in the MSSM, there is one physical CP-odd scalar, $A$, with mass:
\begin{equation}
	m_A^2 = \frac{2b}{\sin 2 \beta}.
\end{equation}
On the other hand, the CP-even scalar sector significantly changes due to the effects of the $\rho$ field. 
In the basis~$(h_d, h_u, \rho)$, the CP-even scalar mass matrix reads as
\begin{widetext}
\begin{equation}
	\mathcal{M}_{even}^2 =
	\begin{pmatrix}
		\cos^2 \beta \left(M_Z^2 + \frac{1}{4} g_X^2 x_{H_u}^2 v^2\right) + \sin^2 \beta \, m_A^2 
		&
		-\frac{1}{2} \sin 2 \beta \left(M_Z^2 + m_A^2 + \frac{1}{4}g_X^2 x_{H_u}^2 v^2\right)
		&
		-\frac{1}{2} g_X x_{H_u} \cos \beta M_X v
	\\
	& & 
	\\
		-\frac{1}{2} \sin 2 \beta \left(M_Z^2 + m_A^2  + \frac{1}{4}g_X^2 x_{H_u}^2 v^2 \right)
		&
		\sin^2 \beta \left(M_Z^2 + \frac{1}{4} g_X^2 x_{H_u}^2 v^2\right) + \cos^2 \beta \, m_A^2
		&
		\frac{1}{2} g_X x_{H_u} \sin \beta M_X v
	\\
	&
	&
	\\
		-\frac{1}{2} g_X x_{H_u} \cos \beta M_X v
		&
		\frac{1}{2} g_X x_{H_u} \sin \beta M_X v
		&
		M_X^2 + \tilde m_\rho^2
	\end{pmatrix}.
\end{equation}

Since $M_X^2 \gg M_Z^2$, one can investigate the behavior of the MSSM Higgs masses when $\rho$ is integrated out:
\\
\begin{equation}
	\mathcal{M'}_{even}^2 =
	\begin{pmatrix}
		\cos^2 \beta
		\left[
			M_Z^2 + \frac{1}{4} g_X^2 x_{H_u}^2 
			\left(
		\frac{v^2 \tilde m_\rho^2 }{M_X^2 + \tilde m_\rho^2}
			\right)
		\right] + \sin^2 \beta \, m_A^2 
		&
		-\frac{1}{2} \sin 2 \beta \left(M_Z^2 + m_A^2 + \frac{1}{4}g_X^2 x_{H_u}^2 
			\left(
				\frac{v^2 \tilde m_\rho^2}{M_X^2 + \tilde m_\rho^2}
			\right)
		\right)
	\\
	
	&
	
	\\
		-\frac{1}{2} \sin 2 \beta \left(M_Z^2 + m_A^2 + \frac{1}{4}g_X^2 x_{H_u}^2 
			\left(
				\frac{v^2 \tilde m_\rho^2}{M_X^2 + \tilde m_\rho^2}
			\right)
		\right)
		&
		\sin^2 \beta
		\left[
			M_Z^2 + \frac{1}{4} g_X^2 x_{H_u}^2 
			\left(
				\frac{v^2 \tilde m_\rho^2}{M_X^2 + \tilde m_\rho^2}
			\right)
		\right] + \cos^2 \beta \, m_A^2
	\end{pmatrix}.
\end{equation}
\end{widetext}
In this limit, $\rho$ has the approximate mass
\begin{equation}
	M_\rho^2 = M_X^2+ \tilde m_\rho^2
\end{equation}

The factor multiplying $\sin^2 \beta$ in the 2-2 element of $\mathcal{M'}_{even}^2$ is an upper bound on the lightest Higgs mass squared. Namely, these terms represent the product of the quartic coupling of the Higgs and the Higgs VEV squared. In the MSSM, this is simply $1/4 \left(g_1^2+g_2^2 \right) v^2$, the $Z$ mass. In this model this is extended by $1/4 g_X^2 x_{H_u}^2 v^2$ times a factor that measures the splitting between the mass of $\rho$ and it's mixing with $h_u$. When $\tilde m_\rho^2 \ll M_X^2$, there is a strong mixing between $h_u$ and $\rho$ that cancels out the positive contribution from the new quartic term. In the opposite limit, when $\tilde m_\rho^2 \gg M_X^2$, this mixing effect decouples and one is left with the positive contribution from the new quartic term. Therefore, for a large $\tilde m_\rho$, the tree-level Higgs mass can be increased beyond the $Z$ mass. Then, the upper bound on the lightest Higgs mass in this model is given by
\begin{equation}
m_h^2 \leq M_Z^2 \cos^2 2 \beta +   \frac{1}{4} g_X^2 x_{H_u}^2 v^2  \frac{\tilde{m}_\rho^2}{M_X^2 + \tilde{m}_\rho^2}  \cos^2 2 \beta.
\end{equation}
This bound is saturated in the decoupling limit, $m_A^2 \gg M_Z^2$.

In order to numerically appreciate the impact of the Stueckelberg Mechanism on the Higgs mass, Fig.~\ref{fig:mh0} displays the lightest Higgs mass versus $\tilde m_\rho$ for different values of $\tan \beta$ and assuming $x_{H_u} = 1$, $M_X=2.5$ TeV, $g_X = 0.4$ and $m_A = 500$ GeV. As an example, with the parameters given in Fig.~\ref{fig:mh0}, $\tan \beta = 10$ and $\tilde m_\rho = 2.5$ TeV, the Higgs mass is increased from about 89 GeV to 97 GeV, an increase of about $9 \%$.

\begin{figure}[t!]
	\includegraphics[scale=0.5]{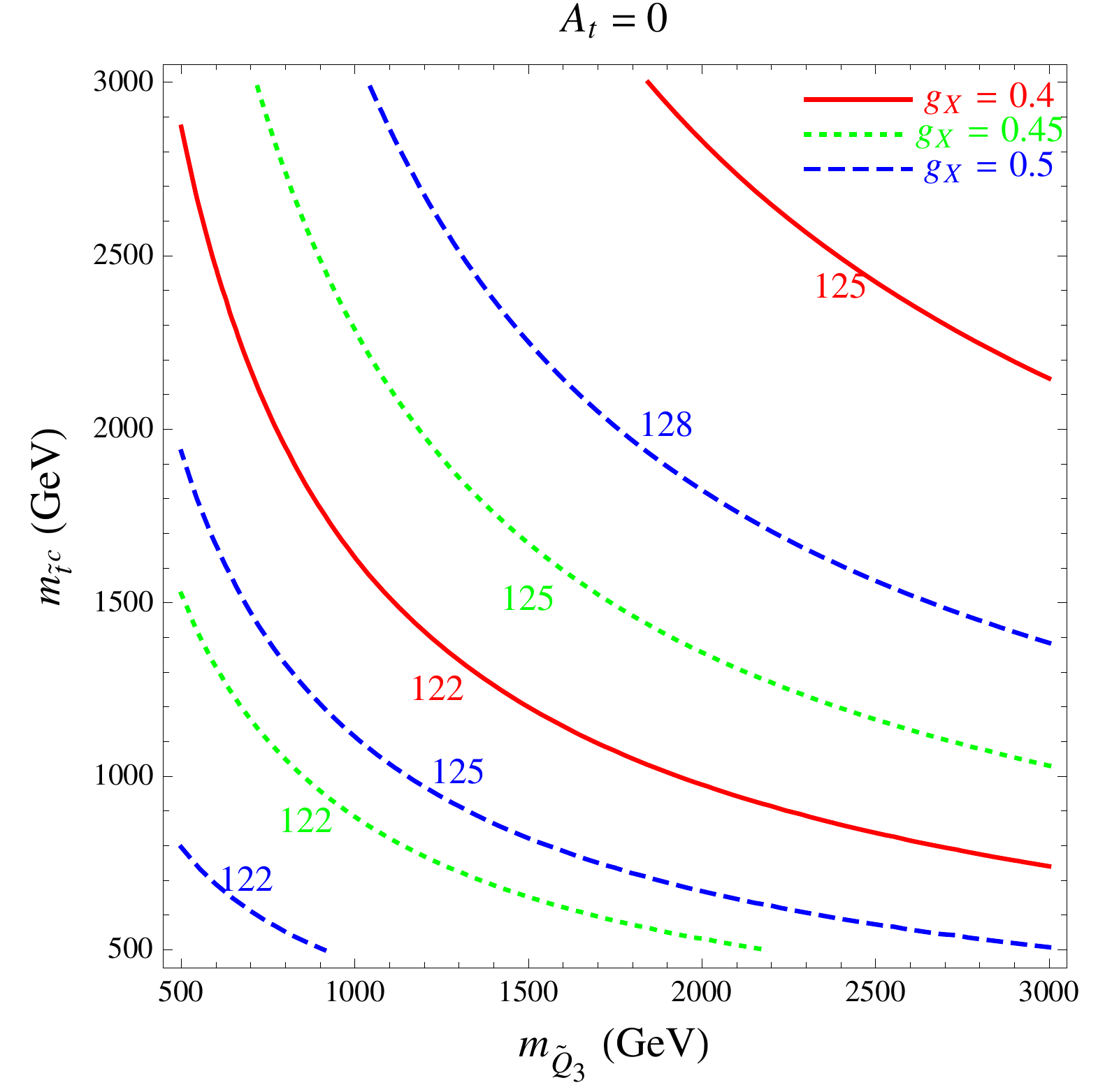}
	\caption{Constant contours of the two-loop level mass of the lightest Higgs, $m_h$ (GeV), in the $m_{\tilde t^c}$ -- $m_{\tilde Q_3}$ plane for $g_X = 0.4$ (red), 0.45 (green and dotted) and 0.5 (blue and dashed), with $m_\rho = 5$ TeV, $M_X = 2.5$ TeV, $\tan \beta = 10$, $x_{H_u} = 1$, $m_A = 500$ GeV, a 3 TeV gluino and $A_t = 0$.}
	\label{fig:mh2}
\end{figure}

\begin{figure}[t!]
	\includegraphics[scale=0.5]{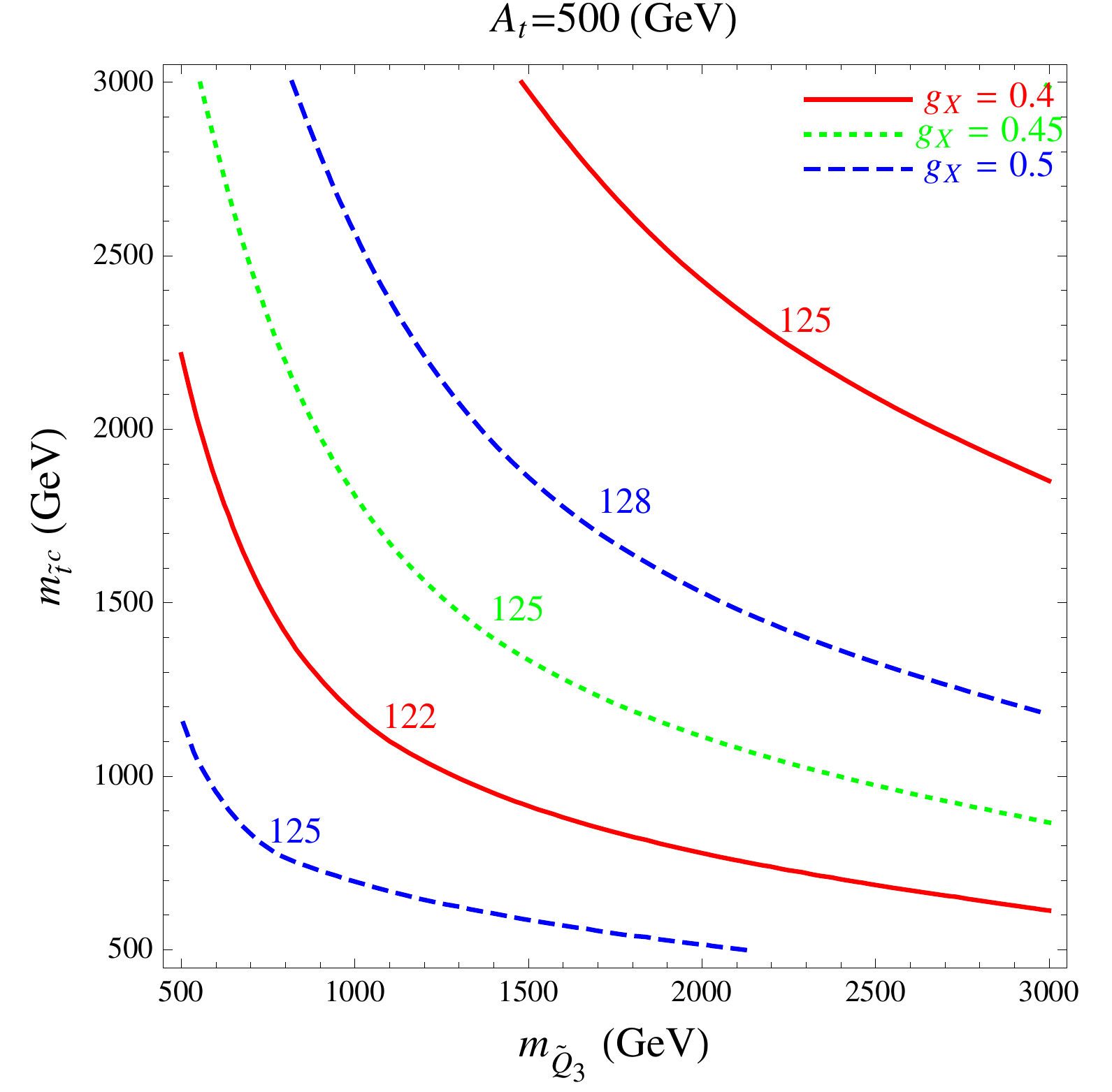}
	\caption{Same as Fig. \ref{fig:mh2} but with $A_t=500$ GeV.}
	\label{fig:mh2b}
\end{figure}

Figures~\ref{fig:mh2} and~\ref{fig:mh2b} extend the calculation of the Higgs mass to the two-loop level\footnote
{
Only the MSSM loop contributions are calculated. These calculations are carried out using FeynHiggs~\cite{Heinemeyer:1998yj}.
}
in order to illustrate that the stops can be light even with a negligible trilinear term. Here we show constant contours of the two-loop level Higgs mass, in GeV, in the $m_{\tilde t^c}$ -- $m_{\tilde Q_3}$ plane for different values of $g_X$ with $\tan \beta = 10$, $\tilde m_\rho = 5$ TeV, a 3 TeV gluino and $A_t = 0$ ($A_t = 500$ GeV )\footnote
{
The $A_t$ term appears in the SUSY breaking Lagrangian as $-y_t A_t \tilde Q_3 H_u \tilde t^c$.
}  in Fig.~\ref{fig:mh2} (Fig.~\ref{fig:mh2b}). As expected, increasing $A_t$ relaxes the constraints on the stop masses. These results are similar to the case where the Higgs mechanism gives mass to the new gauge boson, but our model 
is simpler and guarantees that this symmetry is not broken. 

The unbroken nature of the symmetry has interesting consequences for the conservation of R-parity. This connection can be realized when the $U(1)_X$ is a linear combination of weak hypercharge and $B-L$. In this theory, anomaly cancellation can be achieved by simply adding three generations of right-handed neutrinos, see \textit{e.g.}~\cite{FileviezPerez:2009gr}. Furthermore, it is well-known that this type of symmetry can be obtained from grand unified theories based on the $SO(10)$ gauge group. In this case the R-parity violating terms
\begin{eqnarray}
\hat{L}  \hat{H_u},  \  \,  \hat{Q} \hat{L} \hat{d}^c, \  \, \hat{L} \hat{L} \hat{e}^c, \  {\text{and}}  \  \hat{u}^c \hat{d}^c \hat{d}^c,    
\end{eqnarray} 
are forbidden by the presence of $B-L$ and therefore R-parity is automatically conserved. This guarantees the stability of the lightest supersymmetric particle and that it can be a dark matter candidate in the usual way. See Refs.~\cite{nath1,nath2, Feldman:2011ms} for the study of dark matter candidates in models with the Stueckelberg mechanism. Such a scenario allows Yukawa couplings between the right-handed neutrinos and the SM neutrinos indicating that neutrinos are Dirac fermions.
%
\section{Concluding Remarks}
While consistent with the MSSM, the recently discovered Higgs boson places some tension on the MSSM. This is because the MSSM requires heavy stops or a large stop trilinear term, with somewhat lighter stops, to reproduce the experimentally measured Higgs mass. Various extensions of the MSSM have been proposed to relieve this tension through either new $F$-term or $D$-term contributions to the tree-level Higgs mass. In this letter, we have proposed a new scenario to achieve this goal. Similar  to the case of non-decoupling $D$-terms associated with a new Abelian gauge group, but here, the new gauge boson acquires mass through the Stueckelberg mechanism.

Because the Stueckelberg mechanism allows gauge boson to acquire masses without breaking their associated Abelian symmetry, it can explain the origin of R-parity conservation. This is possible when the new symmetry is related to $B-L$, because R-parity is a subgroup of $B-L$. It this way, the gauge symmetry can play a dual role of both explaining the conservation of R-parity and increasing the Higgs mass at tree level. Since this mechanism provides a solid foundation for R-parity conservation, it motivates the study of dark matter in this context. In particular, the superpartner of the right handed neutrinos can be a good spin zero candidate. We will investigate the experimental constrains for this scenario in the near future. 
\\
\\
{\textit{Acknowledgments:}}
{\small {P.F.P.\ thanks Pran Nath for discussions and the CCPP at New York University for hospitality. 
S. S. is supported in part by the DOE undercontract No. DE-AC02-76-ER-03071 and by the 
NSF under grant No. 1001296.}}



\end{document}